\documentstyle[twocolumn]{article}
\def\kms{ km s$^{-1}$ }
\def\co{$^{12}$CO $(J=1-0)$}

\def\Msun{M_{\odot \hskip-5.2pt \bullet}}

\def\deg{$^\circ$}

\def\c{\vskip 5mm \centerline}
\def\v{\vskip 3mm}
\def\Msun{M_{\odot \hskip-5.2pt \bullet}}
\def\kms{km s$^{-1}$}
\def\r{\hangindent=1pc  \noindent}
\def\deg{$^\circ$}

\def\ha{H${\alpha}$ } 
\def\rc{rotation curve}

\def\pvd{position-velocity diagram}

\begin{document}

\title{The Virgo High-Resolution CO Survey \\
II. Rotation Curves and Dynamical Mass Distributions}

\author{Y. Sofue$^1$,  J. Koda$^{1, 2}$, H. Nakanishi$^1$, S.
\& Onodera$^1$\\
1. Institute of Astronomy, University of Tokyo, Mitaka, Tokyo 181-0015\\
2. Nobeyama Radio Obs., National Astron. Obs., Mitaka, Tokyo 181-8588\\
{\it E-mail sofue@ioa.s.u-tokyo.ac.jp}}

\date{}

\maketitle

\begin{abstract}
Based on a high-resolution CO survey of Virgo spirals  with
the Nobeyama Millimeter-wave Array, we determined the dynamical centers using
velocity fields, and derived position-velocity diagrams (PVDs) along the major
axes of the galaxies across their dynamical centers.
We applied a new iteration method to derive rotation curves (RCs),
which reproduce the observed PVDs.
The obtained high-accuracy RCs generally show steep rise in the central 
$\sim 100$ to 200 pc regions, followed by flat rotation in the disk.
We  applied a deconvolution method to calculate the surface-mass density (SMD)
using the RCs based on two extreme assumptions that the mass distribution is 
either spherical or disk shaped.
Both assumptions gave nearly identical results, agreeing with each other
within a factor of two at any radii.
The SMD distributions revealed central massive cores with the peak SMDs of
$\sim 10^4 - 10^5 \Msun {\rm pc}^{-2}$ and total masses of the order of
$\sim 10^9 \Msun$ within a radius of 200 pc.
Correlation analysis among the derived parameters shows that the central 
CO-line intensities are positively correlated with the central SMD, which
suggests that the deeper is the gravitational potential, the higher is the 
molecular gas density in the nuclei regardless morphological types.

\end{abstract}

\section{Introduction}

%%%
Rotation curves (RC) are one of the most basic tools to discuss the dynamics
of galaxies and to derive the mass distribution. 
Extensive observations have been achieved in radio (HI: Bosma 1981a, b) 
and optical (H$\alpha$: 
Rubin et al. 1980, 1982, 1985, 1998; Mathewson et al. 1993, 1996).
Outer flat RCs were noticed in the early studies
(Rubin et al. 1980, 1982, 1985; Bosma 1981a, b), and are often used to 
estimate the mass distributions in bulges, disks and halos 
(Rubin et al. 1985; Kent 1986; Persic et al. 1996; Takamiya \& Sofue 2000).

Detailed studies of central RCs of spiral galaxies have been 
started only recently (Sofue 1996, 1997; Sofue et al 1999; 
Bertola et al. 1998; Rubin et al 1997, 1999).
In our studies of nearby galaxies, we have shown that the RCs generally rise
steeply in the galaxy centers, and reach rotation velocities  as high
as $\sim 100-300 {\rm km\,s^{-1}}$ within radii of 100 - 200 pc.
On the basis of these results, we have conducted new observations of
the central rotation properties of the Virgo CO rich spirals with 
high accuracy using the Nobeyama Millimeter-wave Array (NMA) in the 
$^{12}$CO($J=1-0$) line (Sofue et al. 2003a; Hereafter Paper I).

In the decades, rotation curves have been derived from position-velocity 
diagrams (PVD) by optical (\ha, NII) and radio (CO, HI) line observations 
(see, e.g., Sofue and Rubin 2001 for a review).
There have been several methods to derive RCs from PVDs.
The traditional method is to trace intensity-weighted velocities
(Warner et al. 1973), and the most popular method is to trace peak-flux 
ridges in PVDs (Rubin et al. 1985; Mathewson et al. 1992).
These methods have given good results for outer regions of galaxies and
for nearly face-on galaxies with sufficiently high spatial resolutions.

Other methods are the terminal-velocity method used for our Galaxy
(Clemens 1985), and the envelope-tracing method (Sofue 1996, 1997; 
Olling 1996), which traces the envelope velocities on PVDs and correct for
the interstellar velocity dispersion and for instrumental resolutions.
This method gives better results for the central regions and for highly
inclined and edge-on galaxies.
We have applied this method to a number of intermediately inclined galaxies to 
obtain central-to-outer RCs (Sofue 1996; Sofue et al. 1999).

These various methods have been the most practical ways to derive RCs 
from PVDs.
However, they do not necessarily give good results for the central regions,
where the PVDs show large intrinsic velocity width and high interstellar 
velocity dispersion.
The central PVDs also suffer strongly from the smearing effect due to 
finite beam sizes and finite slit widths.
In fact, it is known that RCs derived by these methods cannot
reproduce the observed PVDs in the innermost regions of galaxies,
when they are convolved with the observed intensity distributions.
In order to resolve these problems, we have developed a new iteration
method (Takamiya and Sofue 2002), which we describe in detail and apply
to our new data in Section 2.

Not only the methods, but also the properties of the used lines and
observational conditions have limited the accuracy of the central RCs.
Resolutions in HI observations were usually tens of arcseconds and are
insufficient to resolve the central disks, and also the HI gas is often
deficient in the centers.
Optical lines are severely absorbed by the central dusty disks, particularly
in cases of edge-on galaxies.
Optical spectra are severely contaminated by the bright continuum emission
of the bulge and nucleus.
The \ha\ emission is often superposed by a broad absorption line due to
the Balmer wing, and  is often superposed by an extremely broad emission line
due to Seyfert activity.

In this context, we have stressed that the CO line is an ideal tool to
investigate the central kinematics and RCs because of the following
advantages (Sofue 1996, 1997; Sofue et al. 1999):
(1) CO gas is strongly concentrated in the central region.
(2) The CO gas has much smaller intrinsic velocity dispersion than
stars, so that the velocities manifest the rotation.
(3) Even though each molecular cloud is optically thick for the CO line, 
the high-velocity rotation and velocity dispersion among clouds 
make the line practically transparent in so far as the galactic-scale 
dynamics is concerned.
(4) High-angular resolution can be achieved by mm-wave interferometry.
(5) High-velocity resolution can be achieved by radio spectrometers combined
with heterodyne receivers.
(6) High sensitivity can be achieved by large-aperture telescopes and
interferometers. 

The Virgo high-resolution CO survey (ViCS)  has been conducted
in order to obtain a high angular- and spectral-resolution database for
a large number of CO-bright Virgo Cluster spirals in the \co\ line
using the Nobeyama Millimeter-wave Array (NMA)of 
NRO\footnote{NRO (Nobeyama Radio Observatory) is a branch of
the National Astronomical Observatory, operated by the Ministry of
Education, Culture, Sports, Science and Technology, Japan.}.
The major scientific motivation was to investigate detailed
central kinematics of the galaxies, particularly the innermost RCs
from analyses of PVDs across the nuclei, and to derive the central 
mass distributions. 
The advantage to observe the Virgo member galaxies is the identical
distance of 16.1 Mpc based on the Cepheid calibration (Ferrarese et al. 1996).
So, 1 arcsec corresponds to 78 pc, and 1 kpc corresponds to $12''.8$.

The observations were obtained in the course of a long-term project
during the winter seasons from 1999 December through 2002 April in
the AB (longest base lines), C (medium) and D (short base lines) array 
configurations of the NMA. 
The resultant spatial resolutions were 2$''$ to 4$''$, and velocity resolutions
were 10 to 20 \kms. 
The detailed description of the survey together with the observational
parameters and reduction procedure are given in Paper I (Sofue et al. 2003).

In this paper we use the observed PVDs to derive  RCs by applying the 
iteration method, and calculate radial distributions of the surface-mass 
density (SMD) by applying a SMD-deconvolution method described in 
Takamiya and Sofue (2000).

\section{Rotation Curves using the Iteration Method}

\subsection{Iteration Method}

RCs have been derived from observed PVDs, and thus, should be able to
reproduce the original PVDs when convolved  with intensity profiles.
The iteration method (Takamiya \& Sofue 2002) determines a RC
so that it becomes consistent with observations on PVD-basis.
We describe the detailed procedure below. The algorithm is illustrated in 
a flow chart in figure 1. We assume axisymmetry of a galactic disk
throughout the iteration. For a simple explanation, we describe below
an observed PVD as $PV_{\rm obs}$, a model PVD and RC at the $i$-th iteration
as ${PV}_i$ and $V_i$, respectively.

\c{ --- Fig. 1 ---}

Before the iteration, we  define a reference RC ($V_{\rm ref}$) and
a three-dimensional intensity distribution ($I_{\rm ref}$) in a galaxy disk.
$V_{\rm ref}$ is defined as a radial velocity profile traced at the
20\%-level envelope of the peak flux in the $PV_{obs}$.
The intensity, $I_{\rm ref}$, is assumed to have a form:
\begin{equation}
I_{\rm ref}(r,z) = A \Sigma(r) {\rm sech}^2(z/h_z),
\end{equation}
where $\Sigma(r)$ is obtained by integrating $PV_{\rm obs}$ along the velocity
at each radius, ${\rm sech}^2(z/h_z)$ is a model $z$-direction structure,
and $A$ is a normalization constant.
We assume the scale hight of $h_z = 100 $  pc, and the intrinsic
interstellar velocity dispersion is isotropic and is $\sigma_v=10 $ \kms\ 
in the gas disk.
An initial guess of RC ($V_0$) is set to be $V_0=V_{\rm ref}$.
Then the iteration proceeds as follows.

{\bf Step 1:} Constructing a model galaxy by convolving $V_i$ and
$I_{\rm ref}$ with taking $\sigma_v$ into account.

{\bf Step 2:} Virtually observing the model galaxy to obtain ${\rm PV}_i$
in the same conditions (slit-width, resolution, inclination, etc.) that
$PV_{\rm obs}$ was observed.

{\bf Step 3:} Determining a temporal RC ($V_{\rm t}$) by tracing the 20\%-level
envelope of the peak flux in ${\rm PV}_i$.

{\bf Step 4:} Calculating the difference
\begin{equation}
\Delta V_i \equiv V_{\rm ref} - V_{\rm t},
\label{eq:dev}
\end{equation}
which should be close to the difference between $V_i$ and the real RC
that we want to obtain.

{\bf Step 5:} Applying an iteration termination criteria,
\begin{equation}
|\Delta V_i| < \sigma_{\rm tol},
\label{eq:tscrit}
\end{equation}
where $\sigma_{\rm tol}$ is an arbitrary tolerance velocity; we set it as
$\sigma_{\rm tol} = 20 {\rm km\,s^{-1}}$.
If the criterion is not satisfied, we go to Step 6.
If the criterion is satisfied, the iteration terminates, and the 
resultant RC is $V_i$.

{\bf Step 6:} Calculating the $V_{i+1}$ for the next iteration step ($i+1$)
as
\begin{equation}
V_{i+1} \equiv V_i + \Delta V_i,
\end{equation}
and we go back to Step 1.

We made modification to the original method of Takamiya \& Sofue (2002)
in treating  equation (\ref{eq:dev}).
In the original method, only the radial velocity profiles at the 20\%-level
envelope in $PV_{\rm obs}$ and $PV_i$ were considered. We further take
the 20, 40, 60, 80, 100\%-level envelopes ($V_{ref}^l$ and $V_i^l$;
$l=$ 20, 40, 60, 80, 100\%), and replace the equation
(\ref{eq:dev}) with
\begin{equation}
\Delta V = \sum_{l} g_l  \cdot (V_{\rm ref}^l - V_t^l),
\label{eq:dev2}
\end{equation}
where the weights $g_l$ satisfy $\sum_l g_l = 1$.
We usually set $g_1=g_2=...=g_l$, while some $g_l$s are occasionally
set to be zero in cases that those $V^l$ are ill-determined owing to noises.
This modification avoids instability occurred in the iteration that comes
from fluctuations on the 20\%-level envelopes.
When $g_l=1$ for $l=20\%$ and 0 for the others, equations
(\ref{eq:dev}) and (\ref{eq:dev2}) become identical.

\subsection{Center Positions}

The center positions are adopted from Paper I, where
 we  applied the task GAL in the AIPS reduction package,
which fits the velocity field (the first moment map of the RA-Dec-Velocity
cube), by a Brandt rotation curve (Brandt and Scheer 1965).
Table 1 lists the thus obtained dynamical centers and systemic velocities.
In most cases  the observed centers of the fields of view
coincided well with the observed centers within one arcsecond.
However, some deviation was found for NGC 4254 and NGC 4402, and we
reset the centers of these galaxies to the determined dynamical centers.

The GAL fit gave also the inclination and position angle of the inner disk.
However, these quantities are dependent on the streaming motions superposed
on the circular rotation, which are not negligible in the central regions.
Hence, we adopted the inclination and position angles of the whole galaxies
determined from optical images of wider disks.
In table 1 we list the adopted inclination angles taken from
Paper I and the literature therein.
We adopt these values for constructing PVDs and rotation curves.

%%%%%%%%%%%%%% Table 1 Dycent etc. Slit width %%%%%%%%%%%%%%%
\begin{table*}
\begin{center}
\caption{
%%%
{Adopted center positions, systemic velocities, and 
parameters for RC-PVD iteration}
}
%%%
\vskip 2mm
\begin{tabular}{cccccccccc}
\hline\hline
\\
Galaxy & RA (J2000) & Dec. (J2000)  & $V_{\rm sys,lsr}$ & Inclination
& {Beam size} &Slit width for PVD \\
 &  &  & (\kms) & (\deg) & ($''$) & ($''$) \\
\hline
\\
NGC 4192 &  12h13m48.29s &  +14d54m01.9s &  -127.3 &  74.0 &  2.0 &   3.0 \\
NGC 4254 &  12h18m49.61s &  +14d24m59.6s &  2410.8 &  28.0/42.0 &2.5 &3.0 \\
NGC 4303 &  12h19m21.67s &  +04d45m03.7s &  1557.9 &  25.0 &  2.5 &   3.0 \\
NGC 4402 &  12h26m07.44s &  +13d06m44.7s &   216.4 &  75.0 &  2.5 &   3.0 \\
NGC 4419 &  12h26m56.40s &  +15d02m50.2s &  -188.6 &  67.0 &  3.0 &   3.0 \\
\\
NGC 4501 &  12h31m59.12s &  +14d25m13.3s &  2261.1 &  58.0 &  4.5 &   3.0 \\
NGC 4535 &  12h34m20.35s &  +08d11m52.2s &  1966.3 &  43.0 &  3.0 &   3.0 \\
NGC 4536 &  12h34m27.08s &  +02d11m17.0s &  1801.7 &  67.0 &  2.2 &   3.0 \\
NGC 4548 &  12h37m57.47s &  +14d13m18.0s &   464.9 &  37.0 &  2.3 &   3.0 \\
NGC 4569 &  12h36m49.79s &  +13d09m48.7s &  -195.8 &  63.0 &  4.0 &   3.0 \\
\\
NGC 4654 &  12h41m25.59s &  +13d24m03.2s &  1051.4 &  52.0 &  5.0 &   3.0 \\
NGC 4689 &  12h50m15.86s &  +13d29m27.4s &  1614.8 &  30.0 &  5.0 &   3.0 \\
\hline
\end{tabular}\\
\end{center}
\end{table*}
%%%%%%%%%%%%%%%%%%%%%%%

\subsection{Position-Velocity Diagrams}

In order to obtain PVDs of the observed galaxies, we first rotated the
data cube in (RA, Dec, Velocity) to ($X, Y, V$), so that the first axis $X$
represents the distance in the major axis direction, the second axis $Y$
the distance along the  minor axis, and the third axis $V$ the radial velocity.
We then transposed it to a cube with the axes ($X, V, Y$).
Here, each $(X, V)$ plane represents a PVD sliced at a distance $Y$ from
the minor axis.

The PVD along the major axis was obtained as that at $Y=0$.
In order to increase the signal-to-noise ratio, we averaged
several diagrams within $Y=\pm 1''.5$ from the major axis, or those
within a slit of $3''$ wide across the dynamical center.
However, the beam sizes were comparable to or greater 
than the slit width as shown in table 1.
The effective spatial resolutions in the PVDs are, therefore,
(beam width$^2$ + slit width$^2$)$^{1/2}$, which are typically 
a few seconds of arc ($1''= 78$ pc, and 1 kpc = $12''.8$).
The upper diagrams of figure 2 show the thus constructed
observed  PVDs along the major axes across the dynamical centers 
after correcting for the inclinations of galaxy disks.
The intensities are grey-scaled, so that the maximum intensity is in
black, and the minimum in white.

\c{--- Fig. 2 ---}

\subsection{Rotation Curves}

%%%
We applied the iteration method to the observed PVDs, and obtained
rotation curves for 12 galaxies listed in table 1.
Both sides of each PVD along the major axis were fitted independently.
The obtained RCs are shown in the lower diagrams of figure 2,
where we show the reproduced PVDs calculated by convolving the RCs with 
the observed intensity distributions. In most cases, the convolved PVDs 
well mimic the observations.
We then averaged the absolute values of the RCs from both sides for each
galaxy, and obtained rotation curves as a function of the radius.
We combined the thus obtained CO-line RCs with outer optical RCs from 
Rubin et al. (1999).
We also used outer RCs from Sofue et al. (1999) for NGC 4303 and NGC 4569
for larger radii, where both optical and HI data have been combined.
When there are two or more observations from the literature, we took the
average.
Since the inclination angles used in the literature are not necessarily
the same as those used in our analysis, we have corrected them to the same
inclination angles as in table 1.
Figure 3a shows the combined RCs of the  Virgo galaxies, where the
inner CO RCs from the present observations are drawn by thick lines. 
In figure 3b we plot all RCs in one diagram in order to see their general
property.
%%%

The resultant RCs in figure 3 indicate very steep rise in the central few
arcsec regions, or in radii 100 to 200 pc.
The steep rise near the center is remarkable when we  compare with those
from H$\alpha$ and HI observations.
Note, however, that fluctuations of about $\pm 10$ \kms\ superposed on the
RCs are not real, which arise mainly from patchy distributions 
of emission in the PVDs.

\c{--- Fig. 3 --- }

\section{Mass Distributions}

Radial distribution of the surface-mass density (SMD),
$\sigma(R)$, can be calculated directly from a rotation curve 
(Takamiya and Sofue 2000, and the literature therein).
We may assume that the 'true' mass distribution in a real disk galaxy will
be in between two extreme cases: spherical and axisymmetric flat-disk
distributions, as discussed in detail in Takamiya and Sofue (2000).

\subsection{Spherical Mass Distribution}

If we assume that the mass distribution is spherically symmetric,
the SMD ${\sigma}_{\rm S}(R)$ at a radius $R$ is calculated by,
\begin{eqnarray}
{\sigma}_{\rm S}(R) & = & 2 \int\limits_0^{\infty} \rho (r) dz , \\
 & = & \frac{1}{2 \pi} \int\limits_R^{\infty}
\frac{1}{r \sqrt{r^2-R^2}} \frac{dM(r)}{dr}dr .
\end{eqnarray}
Here, $R$, $r$ and $z$ are related by $r=\sqrt{R^2+z^2}$, $\rho(r)$
the spatial density, and $M(r)$ is the mass inside radius $r$ given by
\begin{equation}
M(r)=\frac{r {V(r)}^{2}}{G},
\end{equation}
where $V(r)$ is the rotation velocity at $r$ given by the \rc.

Note that the equation gives underestimated mass at
$R \sim R_{\rm max}$ because of an edge-effect that the integration
is done only up to a maximum radius of observation, $R_{\rm max}$,
instead of infinitely large $R$.
However, this edge effect is negligible for the calculated result
within a radius slightly smaller than $R_{\rm max}$, because the
factor ${(\sqrt{r^2-R^2})}^{-1}$ rapidly diminishes as $r$ gets away
from $R$.

\subsection{Flat-Disk Mass Distribution}

If we assume that the galaxy comprises an infinitely thin flat disk,
the SMD, ${\sigma}_{\rm D}(R)$, can be calculated by
solving the Poisson's equation:
$$
{\sigma}_{\rm D}(R)  =\frac{1}{{\pi}^2 G} \times
$$
$$
\left[ \frac{1}{R} \int\limits_0^R
{\left(\frac{dV^2}{dr} \right)}_x K \left(\frac{x}{R}\right)dx
+ \int\limits_R^{\infty} {\left(\frac{dV^2}{dr} \right)}_x K \left
(\frac{R}{x}\right) \frac{dx}{x} \right], \eqno(9)
$$
where $K$ is the complete elliptic integral and becomes very large
when $x\simeq R$ (Binney \& Tremaine 1987).

When we calculate ${\sigma}_{\rm D}(R)$, we must take into account the
following points.
First, the equation is subject to the boundary condition,
$V(0)=V(\infty)=0$.
Because we have only a few data points for the very central region, where
the assumption of $V(0)=0$ may not apply from the data, the calculated
result near the nucleus would not be reliable.
Hence, in so far as the central $\sim 100$ pc is concerned, the spherical
assumption will give more reliable SMD.
Second, the upper boundary of the integration of the second term is
$R_{\rm max}$, where $V(R)$ is finite, instead of $R=$infinity, where the
equation assumed zero velocity.
This leads to overestimation of  ${\sigma}_{\rm D}(R)$ at
$R\simeq R_{\rm max}$, and the SMD near the outer edges will give an
upper limit.

\subsection{Distributions of the Surface-Mass Density}

Figure 4a shows the thus calculated SMDs for individual galaxies,
where thick dotted lines represent the results for flat-disk assumption, 
and thin full lines for spherical assumption. 
Since the central masses of galaxies are considered to be
dominated by spheroidal components, we may rely more on
the SMDs from spherical assumption.
On the other hand SMDs from flat-disk assumption give better results
for outer regions, where the disk component dominates.
So, in the following, we discuss the SMD's properties in the central
regions inside $r \sim 1$ kpc based on the SMDs from spherical assumption,
and those for outer disks at  $r>\sim 1$ kpc on those from flat-disk
assumption.
In figure figure 4b we show all the SMD profiles from flat-disk assumption
in one diagram in order to see their general property.
Figure 4c shows all SMDs from spherical assumption for the central 2 kpc.

\c{--- Fig. 4 ---}

In spite of the variety of galaxy type and morphology of molecular gas
distributions (Sofue et al. 2003), the SMDs appear to show a basic,
principal structure as the following.

(1) Central massive core: The mass distribution is highly peaked at the
center, showing a core component with scale radius of about
100 to 200 pc. Since the estimated radius is comparable to or smaller than
the beam size, this component is likely not resolved, but may have much
smaller radius.
The calculated central SMDs amount to $\sim 10^4 -10^5 \Msun{\rm pc}^{-2}$,
which is, however,  lower limit, because the core components are not
sufficiently resolved.
The dynamical mass of this component within 100 to 200 pc radius is of the 
order of $\sim 10^9 \Msun$.
In table 2 we list the determined SMDs at the centers and dynamical
masses within radii 200 and 500 pc.

(2) Bulge component: The SMD is then followed by a more gradually decaying
profile at $r=0.2 - 2$ kpc with scale radius of 500 pc.
This component is due to the central bulge.
A typical case of distinct components is seen for NGC 4536.
However, in some galaxies the bulge and core component are not clearly 
distinguishable because of the insufficient resolution.

(3) Disk component: The radial profiles at radii greater at $r\sim 2$
to $\sim 7$ kpc are
approximately expressed by an exponentially decaying function of scale
radius of a few kpc. 

%%%
(4) Massive halo: At radii $r>\sim 7-8$ kpc, the SMD profile becomes
more gently decreasing, indicating an excess over the exponential disk 
due to the massive halo. This is clearly seen in galaxies with RCs 
covering larger radii.
%%%

\subsection{Correlation among dynamical properties}

In table 2 we list the morphological types of the analyzed galaxies,
their total $B$ magnitude taken from NED, and the peak SMD at the center,
$\Sigma_{\rm c}$.
We also list dynamical masses within 200 pc and 500 pc radii from the
dynamical centers, 
$M_{200}$ and $M_{500}$, which were calculated by
$$M_{R}=R V_{\rm rot}^2/G, \eqno(10)$$
where $V_{\rm rot}$ is the rotation velocity given by the RCs, and
$G$ the gravitational constant.

\begin{table*} 
\begin{center}
\caption{Derived mass parameters and CO intensities.}
\begin{tabular}{ccccccccc}
\hline\hline
\\
(1)&(2)&(3)&(4)&(5)&(6)&(7)&(8)
\\
\\
Galaxy & Morphology &$B_{\rm tot}$ &Center SMD 
&$M_{200} $ &$M_{500} $ & $I_{\rm CO,c}~{\rm cos} i $
& $I_{\rm CO, p}~{\rm cos} i $\\
\\
&&[mag] & $[10^4\Msun{\rm pc}^{-2}]$ & $[10^9\Msun]$  & $[10^9\Msun] $
&[K \kms] & [K \kms]  \\
\\
\hline
\\
NGC 4192 & SAB(s)ab & 10.95 & 3.4 & 1.16 & 6.42 & 73 &  96 \\
NGC 4254 & SA(s)c   & 10.44  & 1.4 & 0.32 & 1.64 & 60 &  82 \\
NGC 4303 & SAB(rs)bc& 10.18 & 1.9 & 0.69 & 1.23 & 160 & 395\\
NGC 4402 & Sb       & 12.55 & 0.5 & 0.14 & 1.03 & 5  &  58 \\
NGC 4419 & SB(s)a   & 12.08 & 1.3 & 0.34 & 1.67 & 100 & 229 \\
\\
NGC 4501 & SA(rs)b  & 10.36 & 2.0 & 1.06 & 4.51 & 88  & 111 \\
NGC 4535 & SAB(s)c  & 10.59 & 2.0 & 0.57 & 3.36 & 200 & 275 \\
NGC 4536 & SAB(rs)bc& 11.16 & 3.0 & 1.10 & 3.72 & 171 & 224 \\
NGC 4548 & SBb(rs)  & 10.96 & 4.2 & 1.92 & 6.92 & 38 &  98  \\
NGC 4569 & SAB(rs)ab& 10.26 & 0.9 & 0.15 & 2.31 & 162 & 266 \\
\\
NGC 4654 & SAB(rs)cd& 11.10 & 0.35& 0.06 & 2.06 & 10 & 27 \\
NGC 4689 & SA(rs)bc & 11.60 & 0.75& 2.98 & 0.13 & 12  & 21 \\
\\
\hline
\end{tabular}
\end{center}
%%% 
(1) NGC number; (2) Morphological type from RC3; (3) Total $B$ magnitude
from NASA Extragalactic Database (NED); 
(4) Surface mass density (SMD) at the center;
(5) (6) Dynamical mass calculated by spherical assumption.
$M_{200}$ and $M_{500}$ are masses within 200 and 500 pc radii, 
respectively.
(7) CO intensity at the center calculated by ellipse fitting,
taken from Paper I corrected for inclination.
(8) Peak CO intensities in the observed first-moment maps, taken from Paper I
corrected for inclination.
Note that the CO peaks are near to the centers, but do not 
necessarily coincide with the center positions.
%%%
\end{table*}

{\it SMD vs B magnitude}:
Figure 5a shows a plot of the center values of SMD (column 4 of table 2)
against total B magnitude (column 3).
There is a trend that brighter galaxies have higher SMD.
This is consistent with the fact that brighter galaxies have
brighter bulges, as shown below for the bulge mass and central SMD.

{\it Mass  vs B magnitude}
Figure 5b is a plot of dynamical masses within radii 200 pc
(column 5 of table 2; filled circles) and 500 pc (column 6; diamonds)
against total B magnitude. The trend is the same as for SMD vs B.
 
{\it SMD vs Dynamical mass}:
Figure 5c plots the center SMD (column 4 of table 2) 
against the dynamical masses within radii 200 (column 5; filled circles)
and 500 pc (column 6; diamonds).
There is a tight correlation between the mass and SMD, which are
related as
$$\Sigma_{\rm c} \propto M_{200}^{0.8}, \eqno(11)$$
and 
$$\Sigma_{\rm c} \propto M_{500}. \eqno(12)$$
This correlation is naturally understood, if the core radii
of the central mass concentrations are not largely different among the
galaxies.

Since the rotation curves are nearly flat at radii 0.2 to 1 kpc in
most galaxies, these dynamical masses and center SMD are positively 
correlated with the bulge mass, if the bulge radius is of the order of 
several hundred pc.
Figure 5d plots the dynamical mass within 200 pc radius against that within
500 pc from the same table, which shows a linear correlation.
The correlation between the center SMD and central mass within 200 pc
with the surrounding dynamical mass
would be somehow related to the correlation found for the masses of bulges
and the central black holes (Kormendy and Richstone 1992). 

\c{--- Fig. 5 a,b,c,d ---}

\subsection{Correlation between CO concentration and dynamical properties}

The seventh column of table 2 lists the integrated CO-line intensities
at the centers determined by ellipse fitting, $I_{\rm CO, c}$,
using the observed first moment maps, as taken from Paper I.
The last column lists the peak CO intensities in the maps, 
$I_{\rm CO, p}$.
The CO intensities are corrected for the inclination. 
Note that the CO intensity peaks do not necessarily coincide with the 
dynamical centers, which are, however, located  near to the centers.

{\it CO intensity vs B magnitude}:
Figure 6a plots the central CO intensity (column 7 of table 2;
circles) and peak CO intensity (column 8; triangles) against total 
B magnitude (column 3).
There is a weak correlation between B luminosity and central and
peak CO intensities: brighter galaxies have higher CO intensities.

{\it CO intensity vs SMD}: 
An interesting correlation was found in figure 6b, where the center
CO intensity  (column 7 of table 2; circles) and peak CO intensity
(column 8; triangles) are plotted against the center SMD (column 4).
There is a clear correlation that the
higher is the SMD, the higher is the peak and central CO intensities.
This fact indicates that the deeper is the central gravitational potential,
the more strongly the CO gas is concentrated in the nuclei.

\c{--- Fig. 6 a,b ---}

%%%

\section{Description of Individual Galaxies}

We describe properties of individual galaxies seen in 
their PVDs, rotation curves, and mass distributions. 
More general description of individual galaxies, including CO intensity 
distributions and velocity fields, are presented in Paper I.

\subsection{ NGC 4192}
A bright CO intensity peak is observed at the center, and the galaxy
is classified as a "central/single-peak" type (Paper I).
The rotation characteristics appears normal with
high velocity rotation at $\sim 500$ pc.
The rotation velocity, then, declines to a minimum  at 3-4 kpc according 
to the outer rotation curve taken from the literature (Sofue et al. 1999).
It is then  followed by a flat rotation, which continues to the outer edge.
The SMD shows that the central core and bulge compose a single component,
which is surrounded by an exponential disk.

\subsection{ NGC 4254}
The velocity field shows a regular spider pattern, indicating a circular
rotation of the disk.
The PVD and RC show a sharp rise of rotation velocity in the central few
arcseconds, reaching a shoulder-like step at 100 \kms, and then the velocity
increases gradually to 150 to 200 \kms, depending on the inclination angle.
Note that there are two possibilities of inclination angles, either 28\deg
inferred from molecular gas disk or 42\deg from K-band images
(Sofue et al. 2003b).
The SMD shows clearly a massive core with a sharp peak and a bulge component.
They are surrounded by a disk component, while the disk part is uncertain 
for the limited coverage in the present data.

\subsection{ NGC 4303}
The intensity distribution is classified as twin-peaks, and shows two
prominent off-set bar ridges (Koda et al. 2003).
Due to the non-circular motion, the PVD has a wider velocity dispersion
compared to the other galaxies.
The RC indicates steep rise of rotation within a few arcseconds,
but it  is wavy at $r< 3$ kpc radius. 
Such wavy behavior of RC is typical for a barred galaxy (Sofue et al. 1999).
Accordingly, the SMD behaves peculiar at 1 to 4 kpc, and no firm conclusion
about the inner mass distribution can be obtained.
However, the velocity field in the very central region within radius 300 pc
shows a regular spider pattern, where the steep rise of rotation is observed.
Hence, the central massive component shown in the SMD would be real, as well
as that at $r>5 kpc$.

\subsection{ NGC 4402}
The CO intensity distribution shows a high density nuclear disk
 of radius  $ \sim 10''$, which is surrounded by a more broadly
distributed molecular disk with arm structures.
The PVD and RC shows steeply rising nuclear rotation, followed by a gradually
rising outer disk rotation.
The SMD indicates a sharp peak at the center, suggesting a massive core
with small scale radius, while the present resolution is not sufficient.
It is then surrounded by a bulge and disk components.

\subsection{ NGC 4419}
The CO gas is strongly concentrated in the central $5''$ radius disk,
showing a single-peak surrounded by outer disk.
The central disk has steeply rising rotation, followed by a gradual rise
of the disk component.
The SMD comprises a central core plus bulge component, while the
disk part does not show up because of the insufficient radius coverage. 

\subsection{ NGC 4501}
A compact central peak and spiral arms are prominent in the intensity map.
The velocity field is normal, showing a typical spider pattern.
The PVD shows a central component superposed by a disk/arm component.
The RC indicates a sharp central rise followed by a gradually rising part and
flat outer rotation.
The SMD indicates a central core plus bulge and disk.

\subsection{ NGC 4535}
A high CO gas concentration is found at the center, and the velocity 
field is normal, showing a spider pattern.
The PVD comprises a single ridge, and the RC rises steeply near the
center, reaching a shoulder like maximum followed by a wavy frat part
in the disk.
SMD indicates a core, bulge and disk.

\subsection{ NGC 4536}
This is also a single-CO peak galaxy with a typical spider velocity field.
The PVD comprises two ridges, a central steeply rising component and a
rigid-body like disk component.
The RC rises extremely steeply in the central 100 pc, reaching a maximum
followed by a flat RC.
SMD shows clearly distinguishable three components of the core, bulge 
and disk.

\subsection{NGC 4548}
CO gas is concentrated in the central few arcsec region, and the intensity
is weak.
However, the PVD and RC show very high rotation in the central several
hundred parsecs.
Accordingly, the SMD shows a high-density central core, bulge and disk
components.

\subsection{ NGC 4569}
CO gas is distributed in an elliptical ring of radius 500 pc (Nakanishi
et al. 2003), and the velocity field is complex, having anomalously
high non-circular rotation.
The PVD is complicated with two bright ridges having forbidden velocities.
Our iteration program averaged the observed PVD, and obtained an RC,
which gives only averaged rotation velocities.
The SMD shows a bulge component and disk, while its detail in the
central 1 kpc is not reliable because of the uncertainty in the
RC determination.

\subsection{NGC 4654}
CO distribution is highly distorted and lopsided.
The PVD shows rigid-body ridge, and the RC is also gradually rising
at $r<2 kpc$.
Accordingly, the SMD comprises only disk and bulge components with slight
enhancement due to a possible core component near the center.

\subsection{ NGC 4689}
CO gas is broadly distributed with amorphous morphology, but the
velocity field is normal, showing a spider diagram.
The CO intensity is weak and its distribution is patchy.
PVD has a central ridge and extended flat part.
RC shows a sudden rise in the central $\sim100$ pc to a velocity of 
$\sim 70$ \kms, and then rises gradually to a flat part.
No clear compact component is seen in the SMD, while bulge and disk
components show up.

\section{Discussion}

We have analyzed the high-resolution CO-line survey of Virgo spirals
obtained with the Nobeyama Millimeter-wave Array (Paper I) in order to 
derive dynamical parameters of the galaxies. 
The study of Virgo galaxies provided us with a unique opportunity of using
a very accurate, unambiguous Cepheid distance of 16.1 Mpc from the HST
observations (Ferrarese et al. 1996).

We determined the dynamical centers using the observed CO-line velocity fields
obtained by the high resolution CO survey of CO-rich Virgo spirals.
The high concentration of molecular gas in the centers and high-spatial
and velocity resolutions enabled us to determine the dynamical centers
very accurately.
From the data cube, we constructed position-velocity diagrams
along the major axes across the dynamical centers.

By applying the new iteration method to derive exact rotation curves from
PVDs as proposed by Takamiya and Sofue (2002), we determined rotation curves.
Thus obtained  RCs can well reproduce the observed PVDs by convolving with 
the observed CO intensity distributions.
We combined these RCs with outer disk RCs from the literature.
Except for a few cases, the RCs rises steeply near the center within the
spatial resolution.
Attaining a central peak or a shoulder after the rapid rise, 
the RCs are followed by the gradually varying disk RCs which are in 
general flat till the outer edge.
Some galaxies like NGC 4303 and NGC 4659 with strong non-circular 
velocity field due to a bar showed wavy behavior in the central $\sim 1$ kpc.

We, then, calculated the distributions of surface-mass density SMD as
a function of radius on the two extreme assumptions of spherical symmetry and 
flat-disk using the method developed by Takamiya and Sofue (2000). 
Both the spherical and disk results agreed within a factor of two, while the
spherical assumption give more reliable results for the innermost regions,
where bulge and core component are dominant.
On the other hand, the flat-disk assumption give better results for the 
disk and outer regions where the disk component dominates.
In the present study the halo component was not thoroughly detected,
because we have concentrated in the analysis of the innermost RCs.

The SMDs generally have strong central condensation of mass of
scale radii of 100 to 200 pc, which we call the massive central core. 
The core radii are significantly smaller than those of bulges, whose scale 
radii are usually several hundred parsecs to one kpc. 
Dynamical masses of the massive cores within a radius 200 pc are
of the order of $10^9\Msun$.
Since the present spatial resolutions ($2 - 4 ''$, 160 -300 pc) are 
comparable to the supposed scale radii, we may have not fully
resolved the core components. 
Therefore, the scale radii may still be smaller, and hence,
the center SMD derived by the present analysis would give lower limits.
However, the fact that the center SMD is roughly proportional to
the dynamical mass within 200 pc radius (figure 5) suggests that the scale 
radii of the cores would not be largely scattered among the galaxies.

The present results about the SMD distributions confirm 
our earlier results about massive cores 
(Takamiya and Sofue 2000; Sofue et al. 2001; Koda et al. 2002).
Koda et al. (2002) argued in detail that the high-velocity rotation observed in
the central region of NGC 3079 indeed manifests the 
central massive core with deep gravitational potential.

An interesting positive correlation was found between the CO gas 
concentration in the center and the central SMD. 
The observed galaxies are not  dominated by barred galaxies,
and moreover, the CO concentration is not dependent on galaxy types.
This fact implies that the often mentioned inflow mechanism of molecular gas 
due to bar-driven angular momentum loss is not absolutely necessary.
In fact, our analysis in Paper I showed that the highest CO densities
are found more in centrally single CO peaked galaxies than in twin-peaked
galaxies indicative of bar driven-inflow.
{\it The  $I_{\rm CO}$ - SMD correlation indicates simply that the deeper 
is the central potential, the stronger is the gas concentration.}

This is consistent with our recent results about the extremely high density 
molecular core in NGC 3079, which was argued to be the consequence of a very 
deep potential, where the dense gas is maintained gravitationally stable 
due to high velocity shears, and is suppressed from being fragmented to 
form stars (Sofue et al. 2001; Koda et al. 2002).
The mechanism of molecular gas concentration and its maintenance without
suffering from being exhausted by star formation in a deep gravitational
potential would be an interesting subject for the future.

\vskip 10mm

Acknowledgments:
The observations were performed as a long-term project from 1999 December
till 2002 April at the Nobeyama Radio Observatory of the National
Astronomical Observatories of Japan.
We are indebted to the NRO staff for their help during the
observations. 
We thank Dr. V. Rubin for making their rotation curves available in
tabular format.
JK was financially supported by the
Japan Society for the Promotion of Science (JSPS) for Young Scientists.

%%%%%%%%%%%  References %%%%%%%%%%%%%%%
\vskip 10mm
\parskip=0pt

\noindent{\bf References}
 
\r Bertola, F., Cappellari, M., Funes, J. G., Corsini, E. M., Pizzella, A.,
 Betran, J. C. V., 1998, ApJL 509, 93

\r Binney, J.,  Tremaine, S. 1987, in {\it Galactic Dynamics} 
(Princeton Univ. Press)

\r Bosma A. 1981a. AJ 86, 1825 

\r Bosma A. 1981b. AJ 86, 1791.

\r Brandt, J. C. and Scheer, L. S. 1965 AJ 70, 471. 

\r Clemens, D.P., 1985, ApJ 295, 42

\r Dressler, A., \& Richstone, D. O. 1988, ApJ, 324, 701

\r Ferrarese, L., Freedman, W. L., Hill, R. J., Saha, A.,  Madore, B. F.,
et al. ApJ. 1996, 464, 568

\r Honma, M., Sofue, Y., 1997, PASJ 49, 453  

\r Kent, S. M. 1986, AJ, 91, 1301

\r Koda, K.,  Sofue, Y., Kohno, K., Nakanishi, H., Onodera, S., Okumura, S.
K. and Irwin, Judith A. 2002 ApJ. 573, 105.

\r Koda, K.,  Sofue, Y., Nakanishi, H., Onodera, S. 2003 PASJ in this volume.

\r Kormendy, J.,  Richstone, D. 1992, ApJ, 393, 559 

\r Mathewson DS, Ford VL, Buchhorn M. 1992. ApJ  Suppl. 81, 413

\r Mathewson DS, Ford VL. 1996. ApJ Suppl. 107, 97

\r Olling, R. P., 1996, AJ 112, 457

\r Persic, M., Salucci, P., 1995, ApJS, 99, 501

\r Persic, M., Salucci, P., Stel, F., 1996, MNRAS 281, 27

\r Rubin VC, Ford Jr WK, Thonnard N. 1980. {\it Ap. J.} 238:471

\r Rubin VC, Ford Jr WK, Thonnard N. 1982. {\it Ap. J.} 261:439

\r Rubin, V. C., Burstein, D., Ford, W. K., \& Thonnard, N. 1985, ApJ, 289,
 81

\r Rubin, V. C., Waterman, A. H., \& Kenney, J. D. P. 1999, ApJ, 118, 236

\r Sofue, Y. and Rubin, V. 2001, Ann. Rev. Astron. Astrophys. 39, 137.

\r Sofue, Y., 1996, ApJ 458, 120

\r Sofue, Y., 1997, PASJ 49, 17

\r Sofue, Y., Koda, J., Kohno, K., Okumura, S. K., Honma, M., Kawamura, A.
and Irwin, J. A. 2001 ApJ.L. 547  L115

\r Sofue, Y., Koda, J., Nakanishi, H., Onodera, S., Kohno, K., Okumura, S.K.
, and Tomita, A.  2003a PASJ in this volume (Paper I).

\r Sofue, Y., Koda, J., Nakanishi, H., Hidaka, M. 2003b PASJ in this volume

\r Sofue, Y., Rubin, V.C. 2001 ARAA 39, 137

\r Sofue, Y., Tutui, Y., Honma, M., Tomita, A.,  Takamiya, T.,  Koda, J.,
and Takeda, Y. 1999 ApJ.  523, 136

\r Takamiya, T, \& Sofue, Y., 2000, ApJ 534, 670

\r Takamiya, T, \& Sofue, Y., 2002, ApJ Letters, in press.

\r Warner P.J. Wright M.C.H, Baldwin J.E. 1973, {\it MNRAS}. 163, 163

\newpage

\parindent=0pt
\parskip=4mm
\noindent Figure Captions
\vskip 5mm
PS figures are avaiable from 
http://www.ioa.s.u-tokyo.ac.jp/radio/virgo/
\vskip 5mm
\def\v{\vskip 4mm}

Fig. 1. Algorithm of the iteration method to derive an exact rotation
curve from an observed \pvd\ (Takamiya and Sofue 2002).
\v

Fig. 2. Upper panels: Observed PVDs along the major axes after correcting 
for the inclination of galaxy disks.
Lower panels: Obtained rotation curves by the iteration method, and
reproduced PVDs by convolving the RCs with observed intensity
distributions.
Here, $1''= 78$ pc and 1 kpc =$12''.8$ for the adopted Virgo 
distance of 16.1 Mpc.
\v

Fig. 3: (a) Rotation curves of the Virgo spiral galaxies obtained by the
iteration method, combined with outer RCs from the literature
(Rubin et al. 1999; Sofue et al. 1999), for which
inclination angles are corrected to the same values as in Table 1.
The CO RCs from the present observations are drawn by thick lines.

(b) All rotation curves in one diagram.

Fig. 4. (a) Radial profiles of the SMD (surface-mass density)
of the Virgo spirals.
Full lines are the results from spherical symmetry assumption, which are
more reliable in inner region.
Dashed lines are the result from flat-disk assumption, which are more
reliable for the disk and outer region.

(b) All SMD profiles from flat-disk assumption plotted in one diagram.

(c) All SMD profiels from spherical assumption for the central 2kpc region
plotted in one diagram.

\v

Fig. 5 (a)  SMD at the centers plotted against total B magnitude.

(b)  Dynamical masses within $r=200$ pc (dots) and 500 pc (diamonds) 
against total B magnitude.

(c) SMD against dynamical masses within $r=200$ (dots) and 500 pc
(diamonds).

(d) Dynamical mass within $r=200$ pc against mass within $r=500$ pc.
\v

Fig. 6. (a)  Peak (triangles) and central (circles) 
CO intensities corrected for inclination angle plotted
against total B magnitude.

(b) Peak (triangles) and central (circles) CO intensities against 
the central SMD.

\end{document}